\documentclass[aps,prd,showpacs,twocolumn,
superscriptaddress]{revtex4-1}

\usepackage{graphicx}
\usepackage{dcolumn}
\usepackage{amsmath}
\usepackage{amssymb}
\usepackage{bm}
\usepackage{color}
\usepackage[normalem]{ulem}
\usepackage[dvipsnames]{xcolor}
\usepackage{hyperref}

\hypersetup{
colorlinks=true, % false: boxed links; true: colored links
linkcolor=blue,  % color of internal links
citecolor=cyan,  % color of links to bibliography
}

\begin{document}
\title{
Electromagnetic fields of slowly rotating magnetized compact
stars in conformal gravity
}

\author{Bobur Turimov}
\email{bturimov@astrin.uz}
\affiliation{Ulugh Beg Astronomical Institute, Astronomicheskaya 33,
Tashkent 100052, Uzbekistan }
\author{Bobomurat Ahmedov}
\email{ahmedov@astrin.uz}
\affiliation{Ulugh Beg Astronomical Institute, Astronomicheskaya 33,
Tashkent 100052, Uzbekistan }
\affiliation{National University of Uzbekistan,
Tashkent 100174, Uzbekistan}
\author{Ahmadjon Abdujabbarov}
\email{ahmadjon@astrin.uz}
\affiliation{Center for Field Theory and Particle Physics and Department of Physics, Fudan University, 2005 Songhu Road, 200438 Shanghai, China}
\affiliation{Ulugh Beg Astronomical Institute, Astronomicheskaya 33,
Tashkent 100052, Uzbekistan }
\author{Cosimo Bambi}
\email{bambi@fudan.edu.cn}
\affiliation{Center for Field Theory and Particle Physics and Department of Physics, Fudan University, 2005 Songhu Road, 200438 Shanghai, China}
\affiliation{Theoretical Astrophysics, Eberhard-Karls Universit\"{a}t T\"{u}bingen, Auf der Morgenstelle 10, 72076 T\"{u}bingen, Germany}
\date{\today}
\begin{abstract}
The exact analytical solutions for vacuum electromagnetic fields of slowly rotating
magnetized compact  stars in conformal gravity have been studied.
Taking the realistic dipolar magnetic field configuration for the star, analytical
solutions of the Maxwell equations for the near zone magnetic and the electric
fields exterior to a slowly rotating magnetized relativistic
star in conformal gravity are obtained. In addition, the dipolar electromagnetic
radiation and energy losses from the rotating magnetized compact star in conformal gravity have been  
studied. With the aim to find observational  constraints on the $L$ 
parameter of conformal gravity, the theoretical results for the 
electromagnetic radiation from the rotating magnetized relativistic star in conformal
gravity have been combined with the precise observational data on the radio pulsars 
periods slow down and it is estimated that the upper limit i.e. the maximum value
of the parameter of conformal gravity is less than
$L \lesssim 9.5 \times 10^5\textrm{cm}$ {($L/M \lesssim 5$)}.
\end{abstract}

\maketitle

\section{Introduction}

The end state of the life-cycle evolution of massive stars from  several to $\sim$
hundred Solar masses through a supernova
explosion may form either a neutron star or a black hole. Collapsed black hole
according to the no-hair theorem
does not have intrinsic magnetic field
(see e.g.~\cite{Ginzburg1964,Anderson70,Zeldovich,Price72,Thorne72a}).
On the contrary the formed neutron stars are highly magnetized objects and  one of
the main aims of the modern astrophysics
of compact relativistic stars is to get a clear understanding of
the configuration, structure and evolution of the stellar magnetic field.
The precise measurements of electromagnetic signals from the
radio-pulsars show  that  the  magnetic
fields of compact relativistic stars decrease in strength with
the stellar age and the recycled old neutron stars have weaker
magnetic fields. The strong electromagnetic field affects the
observational data on the high energetic processes
in the vicinity of the compact stars in all electromagnetic radiation spectra.
Observational data of radio-pulsars and soft Gamma ray repeaters (SGR) have shown
that the surface magnetic field
of a typical neutron star is about $10^{12}\ {\rm G}$, while for
magnetars observed as SGRs and anomalous X-ray pulsars (AXP) it may reach the
extreme values as $10^{15}\ {\rm G}$~\cite{Duncan92,Thompson95}.
Therefore, the comparison of the evolution of
magnetic fields and of the rotation spin down observed in neutron
stars with those modeled and theoretically predicted provide a great
challenge and powerful tool to get the constraints on the neutron
star properties in the extreme physics regime and conditions.
Consequently, the continued analysis of the evolution of the stellar magnetic
fields together with the precise measurement of the spin of
relativistic stars at the various evolutionary stages provides an
opportunity to get the constraints on the alternate theories
gravity (ATG)~\footnote{There is big interest to ATGs in order
to explain modern observational data on so-called
dark matter (DM) and dark energy (DE) in the Universe.} in the strong field regime.
The electromagnetic signal detected from radio pulsars is mainly due the to the %
magneto-dipolar radiation from the rotating
magnetized compact star.  The energy loss due to the
electromagnetic radiation causes the spin-down of the rotating
relativistic star~\cite{Thompson2002b,Thompson04,Rezzolla01c,Rezzolla04,Morozova10b,Morozova11,Ahmedov12b,Ahmedov13,Rayimbaev15}.
The structure of the pulsar magnetosphere and related astrophysical processes have %
been widely studied in literature, see e.g.  \cite{Goldreich69,Sturrock71,Mestel71,Ruderman75,Arons79,Muslimov1990,Muslimov1991,Muslimov92,Muslimov97}.

Thus the strong gravitational field regime near relativistic compact stars can play %
a role of laboratory to test general relativity versus other modified or
alternative theories of gravity. Testing gravity theories using the strong field
regime has been performed for X-ray sources from the stellar black hole
candidates~\cite{Cao17,Bambi17a,Bambi17b,Zhang17,Bambi17c,Bambi17d,Ghasemi16,Bambi16,Abdujabbarov10,Abdujabbarov11a,Hakimov17,Hakimov13}. %
The comparison  of the electromagnetic field and radiation of the compact star with %
the pulsar spin down  can also be used to constrain the alternative theories of
gravity~\cite{Turimov17,Hakimov13}.

The impact of strong electromagnetic fields can be observed by other astrophysical %
processes such as gravitational lensing, motion of test particles, and the
electromagnetic spectrum of accretion
discs ~\cite{Novikov73,Shakura73,Page74,Thorne74,Bambi13,Pugliese16}. An analytical %
solution of the exterior electromagnetic field of a rotating magnetized star in the %
Newtonian limit has been found in~\cite{Deutsch1955}. Interior solutions for the
electromagnetic fields of a constant magnetic density star are studied by many
authors, see, for example,~\cite{Ruffini73}.
General relativistic corrections to the electric and magnetic
field structure outside magnetized compact gravitational objects
have been studied in~\cite{Ginzburg1964} and have been further extended by a number %
of authors ~\cite{Anderson70,Petterson74,Wasserman83,
Muslimov92,Muslimov97,Prasanna97,Geppert00,Page00,
Rezzolla01c,Rezzolla04,Kojima04,Rezzolla16,Morozova14,Rezzolla01d,Abdujabbarov14,Abdujabbarov13b,Abdujabbarov13a,Ahmedov11,Morozova14a,Zanotti12,Morozova12,Morozova10,Ahmedov09,Morozova08,Abdikamalov09a,Gralla16,Kojima14,Kojima18,Karas12a}. %
Magnetic fields of spherical compact stars in a braneworld have been studied in~\cite{Ahmedov08}.

In this work we investigate the vacuum
electromagnetic fields of slowly rotating
magnetized compact stars in conformal gravity~proposed in~\cite{Bambi17}. An
example of Lagrangian in this large class of conformally invariant theories of
gravity is
\begin{equation}
{\cal L} = \phi^2 R + 6g^{\mu\nu} \partial_\mu\phi
\partial_\nu \phi\ ,
\end{equation}
where the scalar field $\phi$ (dilaton) is combined with the Ricci scalar.
This Lagrangian is invariant under conformal transformations
\begin{eqnarray}
&& g_{\mu\nu} \to g_{\mu\nu}^* = S g_{\mu\nu} \, , \nonumber\\
&& \phi \to \phi^* = S^{-1/2} \phi \, ,
\end{eqnarray}
where $S = S(x)$ is a function of the spacetime coordinates.

{Since the world around us is not conformally invariant, conformal
symmetry must be broken, and one of the possibilities is that
it is spontaneously broken. In such a case, Nature must select one of the vacua,
namely a certain gauge corresponding a specific choice of the conformal factor $S$. %
In the symmetric phase, the theory is invariant under conformal transformations,
i.e. the physics is independent of the conformal factor $S$. In the broken phase,
the choice of the conformal factor $S$ does lead to observational effects. Such a
choice may look arbitrary, but this is a fundamental feature of any spontaneously
broken symmetry, not just of conformal gravity. In what follows, we will consider
the infinite family of conformal factors found in~\cite{Bambi17} because they have %
the property to solve the singularity problem in the Kerr metric.}

In the paper~\cite{Toshmatov17a}
the quasinormal modes of the scalar fields of a black hole
in conformal gravity have been studied. The energy conditions of a black hole in
conformal gravity have been studied in~\cite{Toshmatov17b}. Conformal invariance
preservation at the quantum level has been discussed in~\cite{Modesto14}.

The present paper is organized as follows.
Sect.~\ref{Formalism} is devoted to the vacuum
electromagnetic fields of a rotating magnetized compact
star in conformal gravity and we derive an exact
analytical solutions
of the
Maxwell equations for the magnetic and the electric
fields of a slowly rotating neutron star in
conformal gravity. In the next Sect~\ref{sec:AA},
we calculate the energy losses from a slowly rotating neutron
star in conformal gravity. In Sect.~\ref{Results}, we obtain
astrophysical constraints on the
value of the parameter of conformal gravity, $L$,
from the comparison of theoretical results on spin down of the rotating star
in conformal gravity  with the current observational data from
radio pulsars period evolution.
Finally, in Sect.~\ref{Summary} we summarize
our obtained results.
Throughout the paper, all physical quantities are denoted with ``$^*$''. We use a
space-like signature $(-,+,+,+)$, a system of units in which $G = c = 1$
and, we restore them when we need to
compare our results with observational data.
Greek indices run from $0$ to $3$, Latin indices from $1$ to $3$.

\section{The vacuum electromagnetic fields of a rotating magnetized compact star in conformal gravity\label{Formalism}}

In this section we briefly discuss the electromagnetic fields in
the spacetime of a magnetized compact star in
conformal gravity. One of the most difficult
mathematical problems is to solve the combined  Einstein-Maxwell equations,
which are coupled nonlinear differential equations, but one can solve
them in the realistic approximation
when the electromagnetic field plays no role into the
space-time geometry in the vicinity of the compact star (see, for example, \cite{Rezzolla01c,Rezzolla04}). 
Assuming that the energy of the  electromagnetic field
is too weak to modify 
the space-time geometry around the compact star, one can
consider the electromagnetic field in the fixed background 
spacetime and investigate the effects
of the background gravitational field
on the electromagnetic field of the slowly rotating relativistic magnetized star
in conformal gravity.

Even the spacetime of  the most rapidly rotating compact (neutron)
stars observed as millisecond pulsars can be approximately described within
the slow rotation limit~\cite{Hartle68}. In Boyer-Lindquist
coordinates ($t, r, \theta, \phi$) the space-time outside the slowly rotating
magnetized star in conformal gravity can be
expressed through the following line element~\cite{Bambi17}
\begin{eqnarray}\label{metric}
ds^{*2} &=& S(r)\left[- N^2\,dt^2+
\frac{1}{N^2} dr^2 +r^2 d\theta^2
\right.
\\\nonumber
&&
\left.
 + r^2
\sin^2\theta d\phi^2 -2 \omega(r)
r^2\sin^2\theta dtd\phi\right]\ ,
\end{eqnarray}
with
$$
N^2(r) = 1-\frac{2M}{r} \ , \quad r\geq R \ ,
$$
where $M$ is the total mass and $R$ is the radius of
the compact star, $\omega(r)=2aM/r^3$ is the angular
velocity of the dragging of inertial frames and
$a$ is the specific angular momentum of the star, which is
defined as $a=J/M$, and $J=I\Omega$ is the total
angular momentum, with the moment of inertia $I$
and the angular velocity $\Omega$
(or the period $P=2\pi/\Omega$ of rotation of the star)
which are very important and precisely measurable quantities/parameters
in observation of pulsars.

The radial function $S(r)$ in Eq.(\ref{metric})
is the scaling factor and, in the slowly rotating limit,
has the form~\cite{Bambi17}
\begin{eqnarray}
S(r) = \left(1 + \frac{L^2}{r^2}\right)^{2(n+1)} ,
\quad n = 1,2,3, ... \ ,
\end{eqnarray}
{where $L$ is a parameter with dimensions of a length and $n$ is an integer
positive number. The theory does not provide any prediction for the value of $L$,
so we can expect that $L$ is either of the order of the Planck length, $L \sim
L_{\rm P}$, or of the order of the gravitational radius of the object, $L\sim M$,
as these are the only two length scales of the system~\cite{Bambi17}. The first
option is realized with the scale already present in the action, while the latter
is with the scale that breaks conformal symmetry on-shell. A priori, both scenarios
are possible and natural. In the present paper, we will consider the second option
with $L\sim M$, as it is the only one with potential astrophysical implications in
compact objects. If $L \sim L_{\rm P}$, modifications of Einstein's gravity would
only show up in high energy/high curvature regimes. The choice of $n$ is related to
the symmetry breaking. As in any spontaneously broken symmetry, we cannot say why
Nature selects a particular vacuum in the class of good vacua. In our work, we
consider the simplest case $n=1$ and we briefly describe how our results change for
larger values of $n$.}

In order to study the electromagnetic properties
of slowly rotating magnetized compact stars
in conformal gravity, one has to find the solutions of the Maxwell equations
in conformal gravity which can be written in similar way as
as it has been done in~\cite{Rezzolla01a,Rezzolla01c}.

\textbf{Stellar Model:}

Before doing any calculation, we list here the
realistic stellar model assumptions.

\begin{itemize}

\item
The magnetic moment of the relativistic star
does not have strong dependence on time due to the high
electrical conductivity of the stellar medium
$\sigma \to \infty$, see e.g., \cite{Rezzolla01a}.

%%%%%%%%%%%%%%%%%%%%%%%%%%%%%%%%%%%%%%%%%%%%%%%%%%%%%%%
\item
In the slowly rotation limit, the linear approximation
of the angular velocities of rotation is very good one that is
${\cal O}(\omega)$ and ${\cal O}(\Omega)$, respectively.

%%%%%%%%%%%%%%%%%%%%%%%%%%%%%%%%%%%%%%%%%%%%%%%%%%%%%%%
\item
It is assumed that the rotating star has a spherical shape in the slowly rotation
approximation and
the deformation due to the stellar rotation is negligible.

\item The medium  is vacuum outside the star.

%%%%%%%%%%%%%%%%%%%%%%%%%%%%%%%%%%%%%%%%%%%%%%%%%%%%%%%
\item
Using the above assumptions we look for the stationary solutions of the Maxwell
equations for the components of the magnetic field of the rotating star
in conformal gravity in the following form~\cite{Rezzolla01c,Rezzolla01a}
\begin{eqnarray}\label{Bra}
\nonumber
B^{\hat r}(r,\theta,\phi,t) &=& F^*(r)
\\
&\times &
\left[\cos\chi\cos\theta + \sin\chi\sin\theta\cos\lambda\right] ,\ \ \ \
\\
\label{Bta}
\nonumber
B^{\hat \theta}(r,\theta,\phi,t) &=& G^*(r)
\\
&\times &
\left[\cos\chi\sin\theta - \sin\chi\cos\theta\cos\lambda\right] ,\ \ \ \
\\\label{Bfi}
\nonumber
B^{\hat \phi}(r,\theta,\phi,t) &=& H^*(r)
\\
&\times &
\sin\chi\sin\lambda, \qquad \lambda = \phi - \Omega t ,\ \ \ \
\end{eqnarray}
where the unknown radial functions $F^*(r)$, $G^*(r)$,
and $H^*(r)$ are responsible for the corrections to the magnetic field
due to conformal
gravity parameters and neutron star's mass and $\chi$ is the inclination angle of
the magnetic field with respect to the stellar rotation axis.

%%%%%%%%%%%%%%%%%%%%%%%%%%%%%%%%%%%%%%%%%%%%%%%%%%%%%%%
\end{itemize}

In the paper~\cite{Rezzolla01c}, such a consideration
has already  been performed in the general
relativistic case and the expressions for the stationary
vacuum electromagnetic fields of a slowly rotating
relativistic star have been clearly shown.
Following to the  techniques
used in the paper~\cite{Rezzolla01c} we look for the
solutions and relations for the electromagnetic fields
of a slowly rotating compact star
in conformal gravity that are distinguished
by the scaling factor $S(r)$ in comparison
with the general relativistic ones.
Then one can simply write them in the following form
\begin{align}
\left(B^{\hat i}, E^{\hat i}\right)_{\rm CG} =
\frac{1}{S}
\left(B^{\hat i}, E^{\hat i}\right)_{\rm GR} \ , \qquad (i = 1,2,3)\ ,
\end{align}
or
\begin{align}
\left(\textbf{B},\textbf{E}\right)_{\rm CG} =
\frac{1}{S}
\left(\textbf{B},\textbf{E}\right)_{\rm GR}\ ,
\end{align}
where the quantities $\textbf{B}$ and $\textbf{E}$
are the magnetic and the electric fields, respectively.

Collecting all the statements which are introduced here,
one can easily find the profile radial functions
$F^*(r)$, $G^*(r)$, and $H^*(r)$
in the expressions (\ref{Bra})-(\ref{Bfi})
for the components of the magnetic field
in the following form (see e.g.,~\cite{Rezzolla01c})
\begin{eqnarray}
\label{F}
&&
F^*(r)= -\frac{3\mu}{4M^3S}
\left[\ln N^2+\frac{2M}{r}\left(1+\frac{M}{r}\right)
\right]\ ,
\\
\label{G}
&&
G^*(r)=H^*(r)=\frac{3\mu N}{4rM^2S}
\left[\frac{r}{M}\ln N^2+\frac{1}{N^2}+1\right]\ , \ \ \ \ \ \
\end{eqnarray}
where $\mu$ is the magnetic dipole
moment of the magnetized slowly rotating compact star.
From the astrophysical point of view, the electric
field of compact stars (pulsars and magnetars) is
at least $V/c$ times weaker than the stellar
magnetic field, where $V$ is the nonrelativistic linear
velocity of the neutron star surface.
Analytical expressions for the
electric field are given in Appendix \ref{EF}.

\begin{figure}[h]
\includegraphics[width=8.5cm,angle=0]{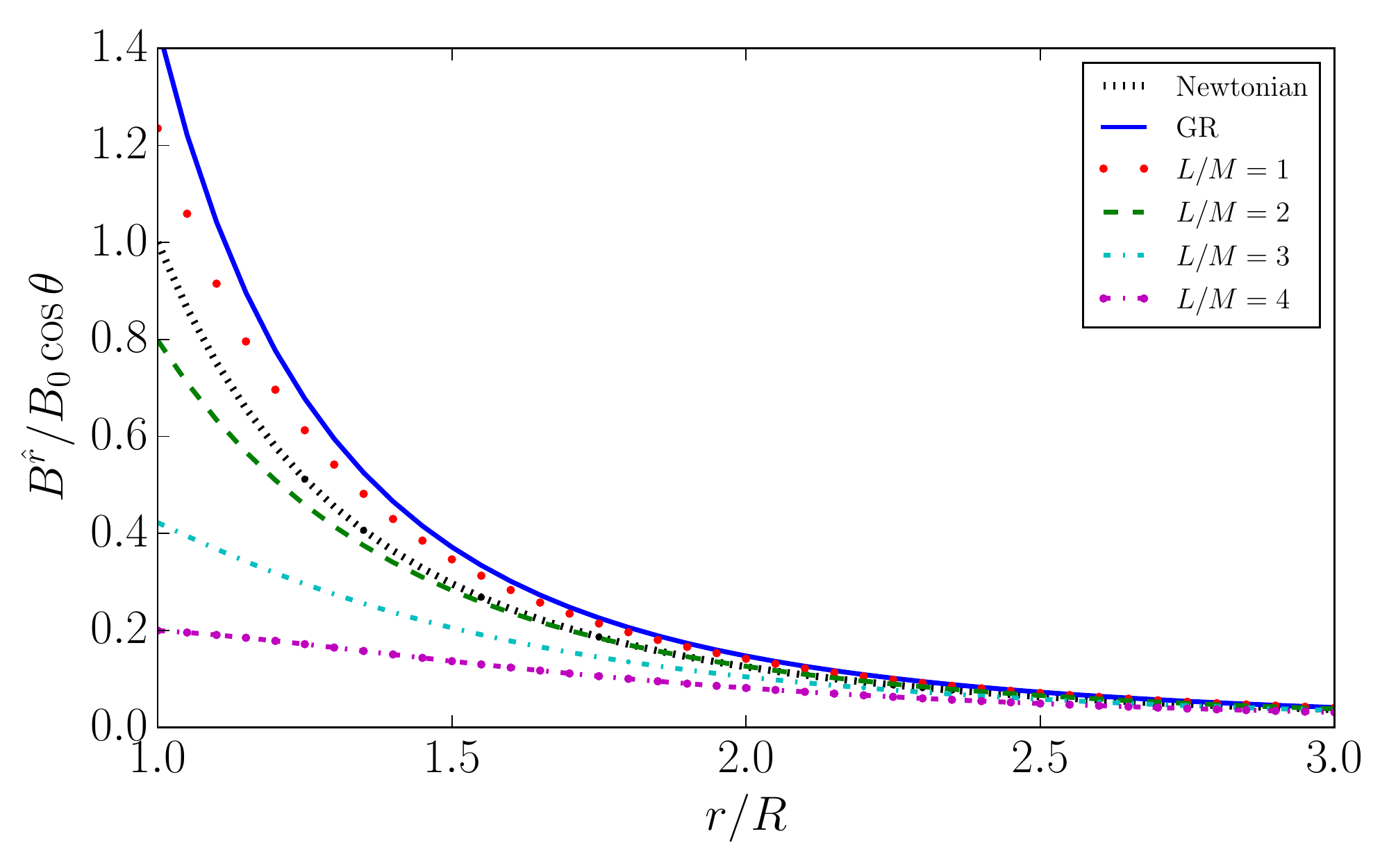}
\caption{ \label{Brad}
Normalized dimensionless radial $r/R$ dependence of the
the radial component of the magnetic field
$B^{\hat r}/B_0\cos\theta$ in conformal gravity
for the compactness $\epsilon =0.3$
with zero inclination angle $\chi = 0$.}
\end{figure}

\begin{figure}[h]
\includegraphics[width=8.5cm,angle=0]{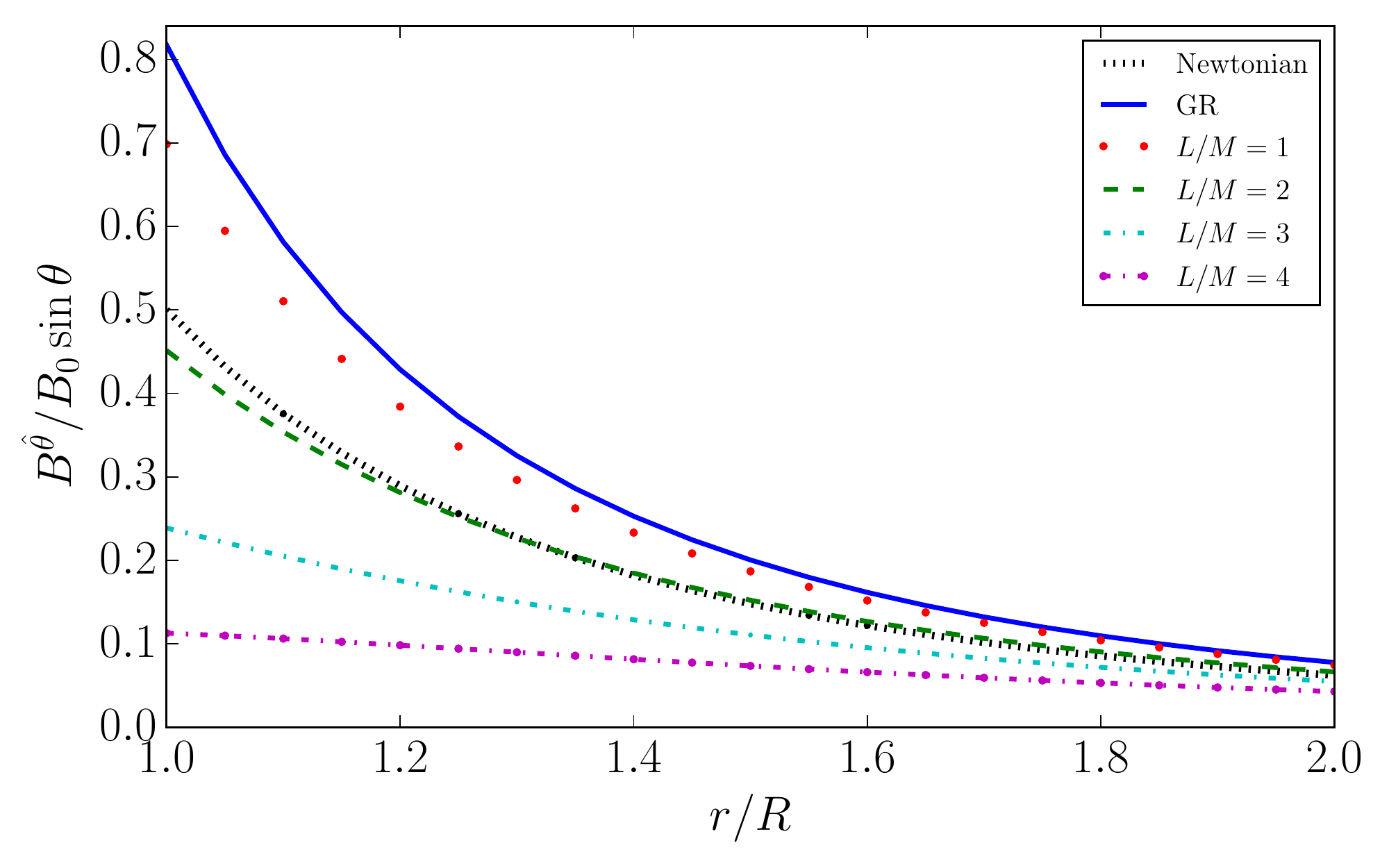}
\caption{ \label{Btan}
The normalized dimensionless radial $r/R$ dependence of the
the tangential component of the magnetic field
$B^{\hat\theta}/B_0\sin\theta$ in conformal gravity
for the compactness $\epsilon =0.3$
with zero inclination angle $\chi = 0$.}
\end{figure}

Hereafter, introducing the normalized dimensionless
radial coordinate $\eta = r/R$ and assuming  zero
inclination angle $\chi = 0$, one can write
the exact solutions for the components of the magnetic field
(\ref{Bra})-(\ref{Bfi}) in the following form
\begin{eqnarray}
\label{Brr}
B^{\hat r}(\eta,\theta) &=& -
\frac{3B_0}{\epsilon^3 S}
\left[\ln N^2
+\frac{\epsilon}{\eta}\left(1+\frac{\epsilon}{2\eta}\right)
\right]\cos\theta \ , \ \
\\
\label{Btt}
B^{\hat \theta}(\eta,\theta) &=& \frac{3B_0N}{
\eta\epsilon^2S}\left[\frac{2\eta}{\epsilon}
\ln N^2 +\frac{1}{N^2}+1\right]
\sin\theta \ , \ \
\\\label{Bff}
B^{\hat \phi}(\eta,\theta) &=& 0\ ,
\end{eqnarray}
where $B_0 = 2\mu/R^3$ is the Newtonian value of the  surface magnetic field
at the polar cap of the star, $\epsilon = 2M/R$
is the compactness of the star and $N^2(\eta) = 1-\epsilon/\eta$ is the
lapse function. The scaling factor can be rewritten in terms
of the normalized dimensionless radial coordinate in the following form
\begin{eqnarray}
S(\eta) = \left[1+\left(\frac{L}{R}\right)^2
\frac{1}{\eta^2}\right]^{2(n+1)}\ , \quad n =1, 2, 3,...\ ,
\end{eqnarray}
and in what follows we will focus on the scenario in which $L$ can be of the order %
of the gravitational radius of the system,
hence also of the order of
the stellar radius $R$. Note that
the scaling factor is always greater than $1$ ($S \ge 1$).
This means, without doing any calculations,
one can conclude that the magnetic field of the
compact star decreases in the conformal gravity.
More precisely, Fig.~\ref{Brad} and Fig.~\ref{Btan}
show the normalized radial dependence
of the radial and the tangential components
of the magnetic field described by the Eqs.~(\ref{Brr}) and (\ref{Btt})
for a relativistic star in conformal gravity  when $n=1$.
One can easily see that for both
components the magnetic field strength are
lowered by increasing the dimensional parameter $L$, which
means the magnetic field of a relativistic star decreases
in the spacetime of rotating relativistic star in conformal gravity.

\section{Astrophysical application \label{sec:AA} }

In this section, we will briefly study
the electromagnetic dipole radiation
from a rotating magnetized neutron star in conformal gravity. Note that such a
phenomenon is at the basis of the observational evidence of radio pulsars
identified with the rotating
magnetized neutron stars. In case of
pure electromagnetic radiation, the luminosity
of the rotating magnetized star in conformal gravity
can be calculated as~\cite{Rezzolla04}
\begin{eqnarray}\label{L*}
L_{\rm em}^* = \frac{\Omega_{R}^{*4}
R^6}{6c^3}B_{\rm R}^{*2}\sin^2\chi\ ,
\end{eqnarray}
where $\Omega_{R}^*$ is the angular velocity
in the observer's frame and $B_{ R}^*$ is the
value of the magnetic field strength at
the surface of the star:
\begin{equation}
\Omega_{ R}^* =  \frac{\Omega}{\sqrt{S_R(1-\epsilon)}}\ ,
\end{equation}
and
\begin{eqnarray}
\label{B*}
B_{ R}^* = B_0\ , \frac{f}{S_R} \ ,
\end{eqnarray}
with
\begin{equation}
f  =  -\frac{3}{\epsilon^3}
\left[\epsilon\left(1+\frac{\epsilon}{2}\right)
+ \ln(1-\epsilon)\right]\ ,
\end{equation}
where the subscript $R$ indicates the value at $r = R$.
{From Eq.(\ref{L*}), one can easily see that
the luminosity of a rotating magnetized
neutron star in conformal gravity is
decreased due to the decrease of the
magnetic field strength and by the gravitational redshift
of the effective rotational angular velocity $\Omega_{\rm R}^*$.}

{In the case of pure dipolar electromagnetic
radiation, the Newtonian value of the luminosity
has the following form~\cite{Landau-Lifshitz2}}
\begin{equation}
L_{0\rm em} = \frac{\Omega^{4} R^6}{6c^3}B_0^{2}
\sin^2\chi\ .
\end{equation}

In order to calculate the rate of the energy loss from the
radio pulsar through dipolar electromagnetic
radiation in conformal gravity, one has to
consider the ratio of the luminosity
in the Newtonian and in conformal gravity~\cite{Rezzolla04}
\begin{equation}
\frac{L_{\rm em}^*}{L_{0\rm em}}
= \left(\frac{f}{1-\epsilon}\right)^2
\left[1+\left(\frac{L}{R}\right)^2\right]^{-8(n+1)} \ .
\end{equation}

\begin{figure}[h]
\includegraphics[width=8.5cm,angle=0]{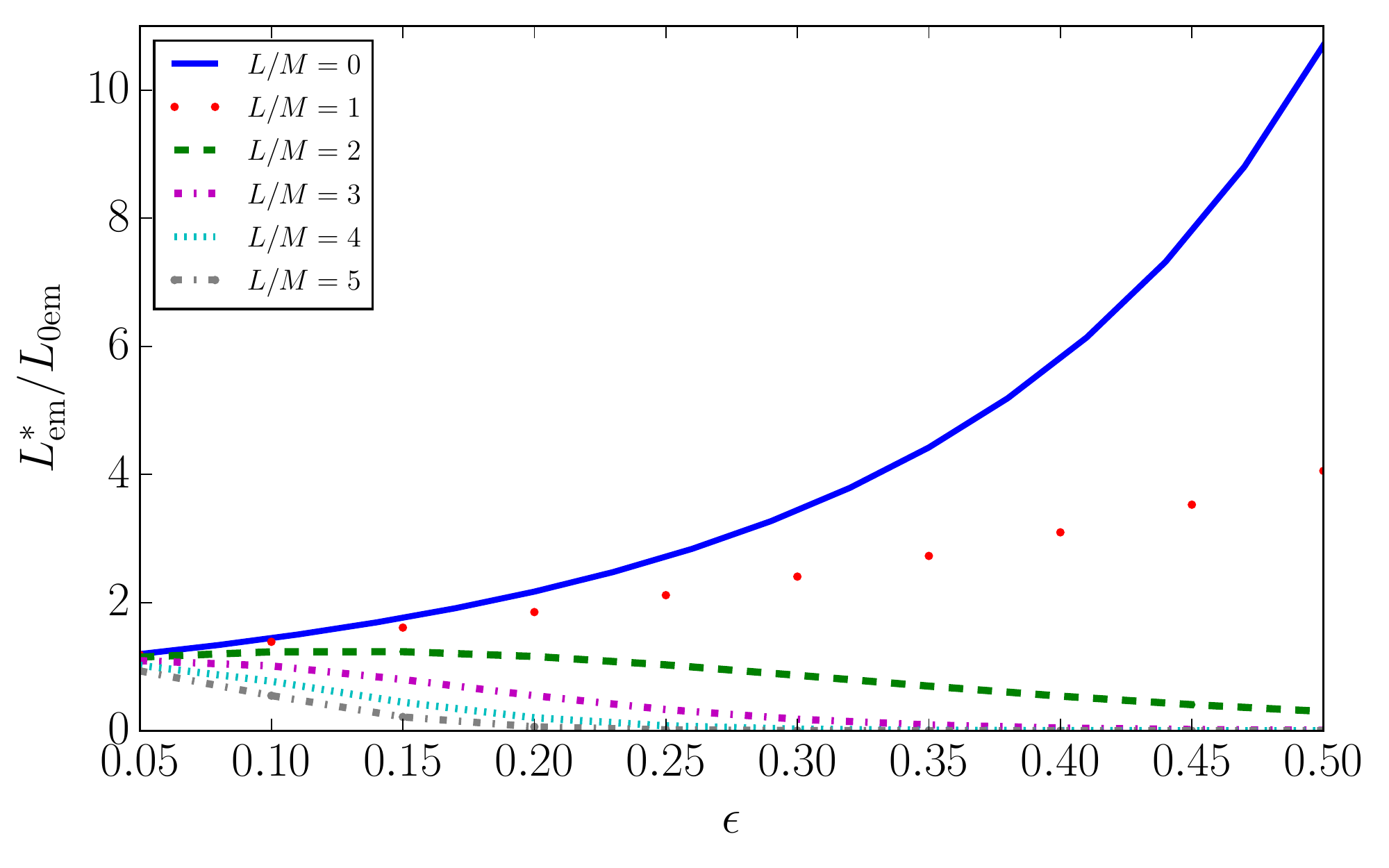}
\caption{ \label{Luminasitye} Dependence of the energy losses
$ L_{\rm em}^*/L_{0\rm em} $ from the compactness
$\epsilon$ of the star for the different values of
the parameter $L/M$.}
\end{figure}
\begin{figure}[h]
\includegraphics[width=8.5cm,angle=0]{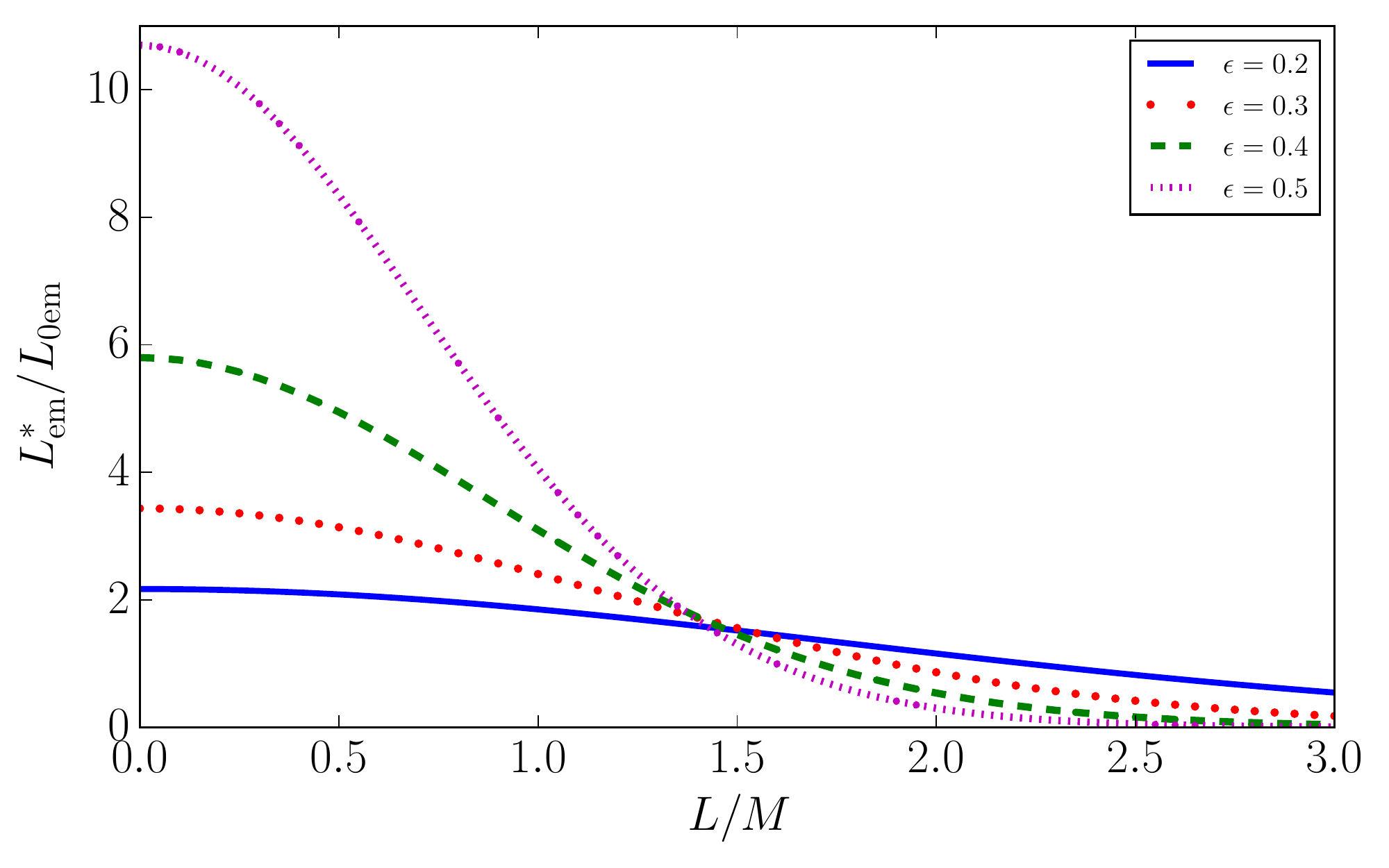}
\caption{\label{LuminasityQ} Dependence of the energy
losses $ L_{\rm em}^*/L_{0 \rm em} $ from the parameter
$L/M$ for the different values of the compactness $\epsilon$ of the star.}
\end{figure}

{The dependence of the rate of energy loss from
the compactness of the magnetized neutron star in
conformal gravity for the different values of the parameter
$L/M$ is illustrated in Fig.~\ref{Luminasitye}. The plot shows
the increase of the rate of energy loss with the increase
of  the compactness of the star.}

In Fig.~\ref{LuminasityQ}  the dependence of
the rate of energy loss of a magnetized neutron
star in conformal from the module of the parameter $L/M$
for the different values of the compactness $\epsilon$ of the star is shown.

\section{Results and discussions \label{Results}}

{Now one can get constraints on
the conformal parameter
$L$ by comparing the obtained theoretical results on electromagnetic radiation
from the rotating magnetized star in conformal gravity with the
observational data on spin down  for the well known
rotating magnetized compact stars and magnetars observed as radio pulsars and SGRs/AXPs.
In order to get upper limit for the
parameter $L$, one can consider the $P-\dot P$
diagram for the typical pulsars~\cite{Guseinov04,Chen93,Johnston17,Ridley10,
Thompson04,Miller15}.
From the observationl data~\cite{Karako15} shown in
Fig.~\ref{PPdot}, one can see that the average value of the
magnetic field strength for a typical radio pulsar is
about $ B_{\rm Av} = B_0 \simeq 10^{12}$~G,
its period is $P\simeq 1$~s, the
period derivative is about $\dot {P}\simeq 10^{-15}$~s~s$^{-1}$,
and the lowest value of the magnetic field strength
in observation is around $ B_R^* = B_{\rm Low}
\simeq 10^{11}$~G (with $P\simeq 1$~s
and $\dot {P}\simeq 10^{-17}$~s~s$^{-1}$).
Using these observational values
and the magnetodipolar formula (\ref{B*})
one can find the upper limit for the value of the
parameter as $L \lesssim 9.5 \times 10^5 \textrm{cm}$
{($L/M \lesssim 5$)} for $n=1$.
This statement is in agreement with the Fig.~\ref{Balpha} and
Fig.~\ref{Bepsilon} on the dependence of the
magnetic field at the surface of the neutron star
from the parameter $L/M$  for the different values
of the compactness of the star.

{
In the table~\ref{Tab1}, dependence of the
model parameters $n$ and $L/M$ is  obtained on comparison of the
the magnetodipolar formula (\ref{B*}) with the observational data on spin down of
the radio pulsars.
\begin{table}
\centering
\caption{\label{Tab1}
Dependence of the parameters $n$ and $L/M$ from comparison of
the magnetodipolar formula (\ref{B*}) with the observational data
for the fixed value of the stellar compactness $\epsilon \simeq 0.4$.}
\label{tab:example_table}
\begin{tabular}{lccccccr} % four columns, alignment for each
		\hline
		n & 1 & 2 & 3 & 5 & 10 & 20 & 100 \\
		\hline
		L/M & 4.87 & 3.74 & 3.15 & 2.49 & 1.79 & 1.28 & 0.58 \\
		\hline
	\end{tabular}
\end{table}
}

\begin{figure}[h]
\includegraphics[width=8cm,angle=0]{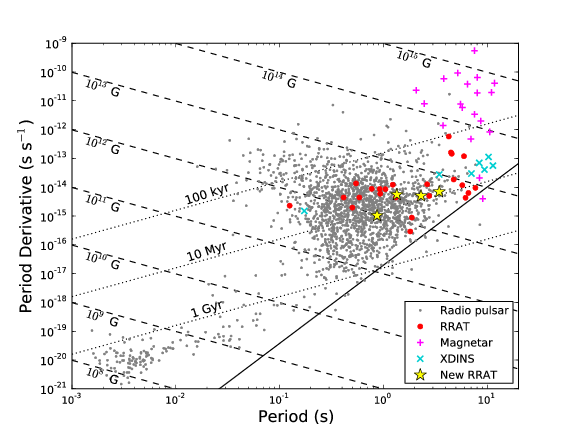}
\caption{\label{PPdot} $P-\dot{P}$
diagram for the observable
pulsars and magnetars from the paper~\cite{Karako15}.}
\end{figure}
\begin{figure}[h]
\includegraphics[width=8cm,angle=0]{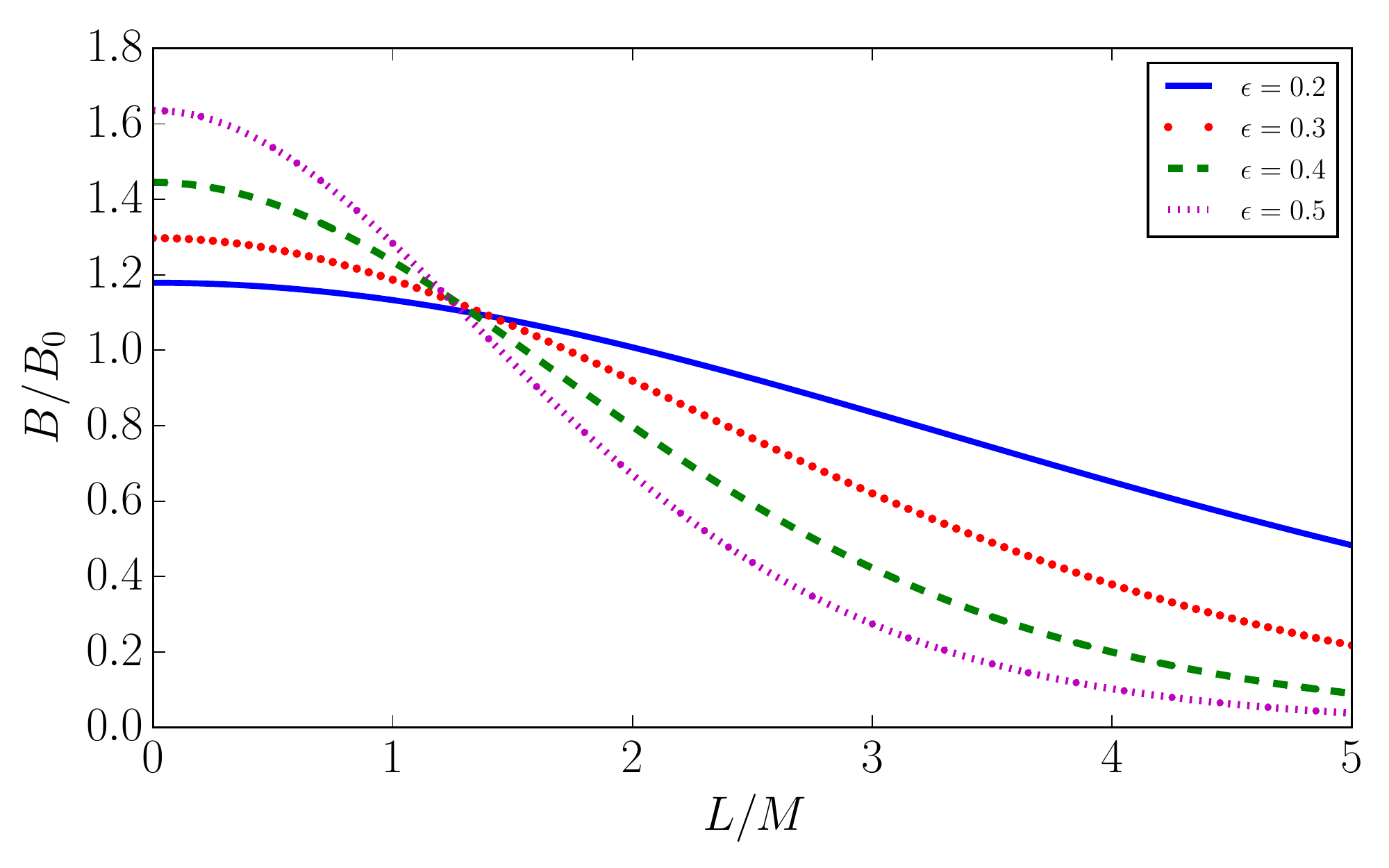}
\caption{ \label{Balpha} Dependence
of the ratio of the magnetic field
from the parameter $ L/M$ for the different
values of the stellar compactness $\epsilon$.}
\end{figure}
\begin{figure}[h]
\includegraphics[width=8cm,angle=0]{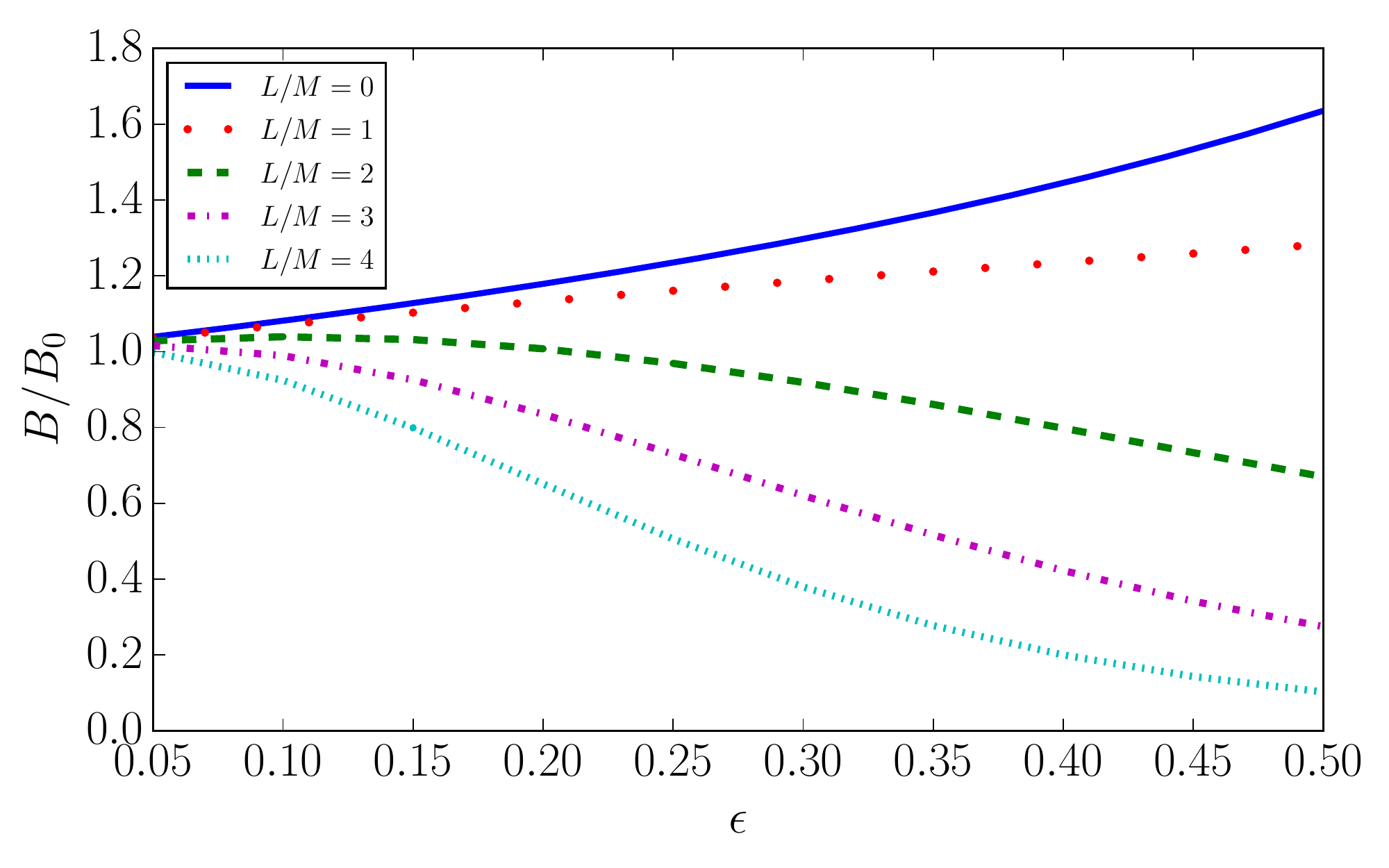}
\caption{\label{Bepsilon} Dependence
of  the normalized magnetic field
from the compactness of the star $ \epsilon$
for the different values of the parameter $L/M$.}
\end{figure}

\section{Summary}\label{Summary}

In the present work we have investigated
the modifications of the electromagnetic fields
of a rotating magnetized compact star
arising from the parameters of the conformal gravity and
their astrophysical implications
to the neutron stars observed as pulsars.
We have studied the  relativistic
Maxwell equations for the dipolar electromagnetic
fields of a slowly rotating magnetized compact star
in terms of the parameters of the conformal gravity and then
{obtained the analytical solutions
for the dipolar magnetic field in terms of
the parameter $L$.}  Along with the magnetic field,
we have obtained the analytical expressions
for the electric field of a
rotating magnetized star in conformal gravity.

As an important application of the obtained results,
we have calculated energy losses of slowly rotating
magnetized neutron star in conformal gravity
through magneto-dipolar radiation and found that
the rotating star with non-zero $L$ parameter
will lose less energy when compared to a
rotating neutron star in
general relativity.
This permits us to check
the effects of the scaling factor arising from the conformal gravity
in the vicinity of a rotating
magnetized star, especially, when one calculates
the electromagnetic luminosity from the star. The latter
is a very important measurable quantity in
pulsar astrophysics.

The obtained dependence from the scaling factor has been combined
with the  astrophysical data on precise measurement of pulsar period slowdown in
order to constrain  the $L$ parameter. We have found
the upper limit for the parameter of conformal gravity
as $ L \lesssim 9.5 \times 10^5$~cm.

\section{acknowledgement}

This research is supported by
by Grants No.~VA-FA-F-2-008 and
No. YFA-Ftech-2018-8 of the Uzbekistan Ministry for Innovation
Development, and by the
Abdus Salam International Centre for Theoretical
Physics through Grant No.~OEA-NT-01.
This research is partially supported by an
Erasmus+ exchange grant between SU and NUUz.
C.B. acknowledges support from the National Natural Science Foundation of China
(Grant No.~U1531117), Fudan University (Grant No.~IDH1512060), and
the Alexander von Humboldt Foundation. B.A. would like to acknowledge
Nazarbayev University, Astana,Kazakhstan for the warm hospitality
through support from ORAU grant SST~2015021.

\begin{appendix}
\begin{widetext}
\section{Electric field of a rotating compact magnetized star in conformal gravity \label{EF}}

The components of the
electric field of slowly rotating
magnetized neutron star in conformal gravity can be chosen
in the following form~\cite{Rezzolla01c}
\begin{eqnarray}
\label{Er}
E^{\hat r}(r,\theta,\phi,t) &=&
\left[f^*(r)+f_3^*(r)\right]
\cos\chi(3\cos^2\theta-1)
%\\\nonumber
%&+&
+3\left[g_1^*(r)+g_3^*(r)\right]
\sin\chi\sin\theta\cos\theta\cos\lambda\ ,
\\\label{Et}
E^{\hat \theta}(r,\theta,\phi,t) &=&
\left[f_2^*(r)+f_4^*(r)\right]
\cos\chi\sin\theta\cos\theta
%\\\nonumber
%&+&
+\left[g_2^*(r)+g_4^*(r)\right]\sin\chi\cos\lambda
%\\\nonumber
%&-&
-
\left[g_5^*(r)+g_6^*(r)\right]
\cos 2\theta\sin\chi\cos\lambda\ , \ \
\\\label{Ef}
E^{\hat \phi}(r,\theta,\phi,t) &=&
\left[g_5^*(r)+g_6^*(r)\right]
\sin\chi\cos\theta\sin\lambda
%\\\nonumber
%&-&
-\left[g_2^*(r)+g_4^*(r)\right]
\sin\chi\cos\theta\sin\lambda\ ,
\end{eqnarray}
where $\{ f_i^*(r) \}$ and $\{ g_i^*(r) \}$
are the functions of the radial coordinate $r$ to be found as solutions
of the Maxwell equations in spacetime of slowly rotating
magnetized neutron star in conformal gravity.

The explicit form of the profile radial  functions is given as
solutions of the Maxwell equations in spacetime of slowly rotating
magnetized neutron star in conformal gravity~\cite{Rezzolla01c}
\begin{eqnarray}\label{f1}
f_1^*(r) &=& \frac{\mu\Omega C^*C_1^*}{6cR^2S}
\left[\frac{2M^2}{3r^2}+\frac{2M}{r}-4
+\left(3-\frac{2r}{M}\right)
\ln N^2\right]\ ,
\\
\label{f2}
f_2^*(r) &=& -\frac{\mu\Omega C^*C_1^*}{cR^2S}
N\left[
\left(1-\frac{r}{M}\right)
\ln N^2
-2 - \frac{2M^2}{3r^2N^2}\right], \ \
\\
\label{f3}
f_3^*(r) &=& {\frac{15\mu\omega r^3}{16cM^5S}}
\left\{
C_3^*\left[\frac{2M^2}{3r^2}+\frac{2M}{r}-4
+\left(3-\frac{2r}{M}\right)
\ln N^2\right]
+
\frac{2M^2}{5r^2}\ln N^2 + \frac{2M^3}{5r^3}\right\}\ ,
\\
\label{f4}
f_4^*(r) &=& {- \frac{45\mu\omega r^3}{8cM^5S}N}
\left\{C_3^*\left[\left(1-\frac{r}{M}\right)
\ln N^2-2 - \frac{2M^2}{3r^2N^2}\right]
+
\frac{M^4}{15r^4N^2}
\right\}\ ,
\\
\label{g2}
g_2^*(r) &=& \frac{3\mu\Omega r}{8cM^3NS}
\left[\ln N^2+\frac{2 M}{r}\left(1+\frac{M}{r}\right)\right]\ ,
\\
\label{g4}
g_4^*(r)&=& -\frac{3\mu\omega r}{8cM^3NS}
\left[\ln N^2+\frac{2M}{r}\left(1+\frac{M}{r}\right)\right]  \ ,
\end{eqnarray}
\end{widetext}
where the constants of integration $C^*$,
$C_1^*$, and $C_3^*$ can be
found from the boundary conditions for the continuity of the
tangential components of the electric field and for the jump of the normal
component of the electric field due to the surface
electric charges at stellar surface.

The other radial functions have the following form as in~\cite{Rezzolla01c}
$$
g_1^*=f_1^*, \qquad g_3^* = f_3^*, \qquad g_5^*=\frac{1}{2}f_2^*,
\qquad g_6^* =\frac{1}{2}f_4^* \ .
$$

\end{appendix}

\bibliographystyle{apsrev4-1}  %% BibTeX style
\bibliography{gravreferences}

\end{document}